\documentclass[12pt,preprint]{aastex}

\def\ga{\mathrel{\raise0.35ex\hbox{$\scriptstyle >$}\kern-0.6em
\lower0.40ex\hbox{{$\scriptstyle \sim$}}}}
\def\la{\mathrel{\raise0.35ex\hbox{$\scriptstyle <$}\kern-0.6em
\lower0.40ex\hbox{{$\scriptstyle \sim$}}}}
\def\co{CO {\it J}=1-0 }
\def\cotwo{CO {\it J}=2-1 }
\def\cothree{CO {\it J}=3-2 }

\def\hij{high-{\it J} }
\def\loj{low-{\it J} }

\newcommand{\kms}{km~s$^{-1}$}

\slugcomment{Accepted for publication in ApJ Letters}

\shorttitle{CO in Himiko}
\shortauthors{Wagg and Kanekar}

\begin{document}
   \title{A deep search for CO~{\it J}=2-1 emission from a Lyman-$\alpha$ blob at $z \sim 6.595$}

   \author{Jeff Wagg\altaffilmark{1} 
               and  Nissim Kanekar\altaffilmark{2} }

\email{jwagg@eso.org}
\altaffiltext{1}{European Southern Observatory, Casilla 19001, Santiago, Chile}
\altaffiltext{2}{Ramanujan Fellow; National Centre for Radio Astrophysics, TIFR, Ganeshkhind, Pune-411007, India}
 
\begin{abstract}
\noindent
We have used the Green Bank Telescope to carry out a deep search for redshifted 
\cotwo\ line emission from an extended  ($>$17~kpc) Ly$\alpha$ blob (LAB), ``Himiko'', 
at $z \sim 6.595$. Our non-detection of \cotwo\ emission places the strong $3\sigma$ 
upper limit of $L^\prime_{\rm CO}  < 1.8 \times 10^{10} \times (\Delta V/250)^{1/2}$~
K~\kms~pc$^2$ on the CO line luminosity. This is comparable to the best current limits 
on the CO line luminosity in LABs at $z \sim 3$ and lower-luminosity Lyman-$\alpha$ 
emitters (LAEs) at $z \gtrsim 6.5$. High-$z$ LABs appear to have lower CO line 
luminosities than the host galaxies of luminous quasars and sub-mm galaxies at similar
redshifts, despite their high stellar mass. Although the CO-to-H$_2$ conversion factor is uncertain for galaxies in the early Universe, we assume $X_{CO} = 0.8$~M$_\odot$~(K~km~s$^{-1}$~pc$^2$)$^{-1}$ to 
 obtain the limit M(H$_2$)~$< 1.4 \times 
10^{10}$~M$_\odot$ on Himiko's molecular gas mass; this is a factor of $\gtrsim 2.5$ 
lower than the stellar mass in the $z \sim 6.595$ LAB.
\end{abstract}


%

\section{Introduction}

Significant populations of star-forming galaxies have recently been discovered at 
high redshifts, $z \gtrsim 6$, towards the end of the reionization epoch 
\citep[e.g.][]{ellis08}. Many of these galaxies have been identified due to their 
excess emission in narrow-band images centred on the redshifted Ly$\alpha$ wavelength 
\citep[e.g.][]{rhoads00,taniguchi05,hu10}; such systems are referred to as Ly$\alpha$ 
emitters (hereafter LAEs). The star-formation rates determined from measurements 
of the rest-frame UV continuum emission in LAEs at $z \sim 6.5$ tend to be low,
$SFR \sim 5 - 10$~M$_{\odot}$~yr$^{-1}$ \citep[e.g.][]{taniguchi05}, and the 
size of the emission region is typically small, $\sim 1$~kpc \citep{cowie11}. 

High-$z$ LAEs thus appear to be relatively small galaxies, probably undergoing 
quiescent star formation. However, a recent Subaru narrow-band imaging survey 
has discovered a highly extended LAE at $z \sim 6.6$, with a spatial extent 
of $> 10$~kpc, far larger than that of the bulk of the $z \sim 6.5$ LAE population
\citep{ouchi09}. The object, ``Himiko'', also has a far higher Ly$\alpha$ 
line luminosity,  $>10^{43}$~erg~s$^{-1}$, than any other LAE at $z > 6$ .
Both the large size and the high Lyman-$\alpha$ luminosity make Himiko similar
to the ``Lyman-$\alpha$ blobs'' (LABs), that have so far been detected at 
$z \sim 2.5 - 4$ (e.g. \citealp{steidel00,matsuda04,dey05,nilsson06,matsuda11}). 
The LABs discovered at $z \sim 3$ are extremely large ($\gtrsim 50$~kpc in size) 
Ly$\alpha$-emitting nebulae that show a striking correlation with over-dense regions 
in the Universe. These have Ly$\alpha$ luminosities $> 10^{43}$~erg~s$^{-1}$, 
similar to the luminosities seen in massive high-$z$ radio galaxies (e.g. 
\citealp{reuland03}, but do not show radio emission. Follow-up multi-wavelength 
studies have shown that LABs tend to be associated with bright sub-millimeter or 
infrared-luminous galaxies (e.g. \citealp{geach05}) or obscured active galactic 
nuclei (AGNs; e.g. \citealp{basuzych04,geach09}), often with high star formation 
rates ($\sim 1000$~M$_\odot$~per year). Recent {\it Spitzer} studies have found
that the infrared (IR) images of 60\% of $z \sim 3$ LABs are consistent with an 
origin due to star formation, with the remaining 40\% of the sample likely to 
arise either due to AGN activity or an extreme starburst \citep{colbert11}. A few 
LABs have also been detected without a known associated bright galaxy (e.g. 
\citealp{nilsson06}). 

No LABs have so far been detected at low redshifts, $z < 1$ \citep{keel09}; the 
blobs thus appear to be a high-redshift phenomenon. The fact that LABs are typically
located in over-dense regions suggests that they are linked to the formation of 
the massive galaxies (e.g. \citealp{steidel00,erb11}). Although a number of mechanisms 
have been suggested for the sources that power the Ly$\alpha$ emission (e.g. cooling 
inflows of gas, outflows of gas from starburst galaxies or AGNs, or even AGN 
photoionization; \citealp{taniguchi00,haiman01,matsuda04,dey05,dijkstra09,geach09}),
it appears unclear whether any single mechanism is capable of explaining the entire 
LAB population. However, while cooling radiation from gas streams may contribute 
some fraction of the LAB power \citep{haiman00,goerdt10}, radiative
transfer calculations have shown that such streams of halo gas cannot account for 
all the Ly$\alpha$ luminosity \citep{fauchergiguere10}. Galaxy formation must hence 
be responsible for a significant fraction of the Ly$\alpha$ emission from LABs. 
Studies of the star formation activity of galaxies in the vicinity of LABs are thus
essential to understanding their nature. Observations of molecular and atomic gas 
tracers in the LABs would provide crucial insight into the fuel for star formation. 
The most effective means of studying the cold molecular gas reservoir in high-$z$ 
galaxies is through observations of redshifted CO line emission (e.g. 
\citealp{solomon05}), which provide a means of studying gas kinematics and 
estimating the total molecular gas mass available for star formation. 

At present, only two LABs have been searched for CO line emission, both at $z \sim 3$,
but with no detections of molecular gas \citep{yang12}. We present here a search 
for \cotwo\ line emission in the $z \sim 6.595$ LAB, Himiko, in order to measure 
its cold molecular gas mass and constrain the kinematics of the interstellar 
gas in a massive galaxy during the epoch of reionization.\footnote{When required, 
this paper uses a $\Lambda$CDM cosmology, with $H_0 = 71$~km~s$^{-1}$~Mpc$^{-1}$, 
$\Omega_\Lambda = 0.73$, $\Omega_m = 0.27$ \citep{spergel07}.}

\section{Himiko: a Ly$\alpha$ blob at z=6.595}


The target of our CO line observations, ``Himiko'', is a giant ($\gtrsim 17$~kpc) 
LAE at $z=6.595$, discovered in the Subaru/\textit{XMM-Newton} Deep Survey 
\citep{ouchi09}. Himiko has an extremely high Ly$\alpha$ line luminosity, 
L(Ly$\alpha$)$=(3.9\pm 0.2)\times 10^{43}$~erg~s$^{-1}$, well into the LAB 
category. The Ly$\alpha$ emission yields a star-formation rate (SFR) of 
$\sim 36$~M$_\odot$/yr \citep{ouchi09};  However, this is likely to be an
underestimate by a factor of at least a few, given that the Ly$\alpha$ line 
is both resonantly scattered by dust, and, for these redshifts, attenuated due 
to absorption by the damping wing of neutral hydrogen in the intergalactic medium. 
The SFR of Himiko is thus likely to be significantly larger than the estimate of 
$36$~M$_\odot$/yr. \citet{ouchi09} used fits of stellar synthesis models to the 
spectral energy distribution over the optical to near-infrared (near-IR) wavelength
range to obtain an extremely high stellar mass, $(0.9 - 5) \times 10^{10}$~M$_\odot$, 
indicating that Himiko is a massive galaxy. The large transverse size of 
the Ly$\alpha$-emitting region is also consistent with Himiko being an extremely 
massive object, in the context of hierarchical models of structure formation. 
It is thus an excellent candidate for a search for the molecular gas that fuels its 
high rate of star formation.


\section{Observations and Data Analysis}

Observations of the \cotwo\ line emission in Himiko were made with the 110m Green 
Bank Telescope (GBT) in project AGBT10B-010, using the Ka-band receiver which covers 
the frequency range $26-40$~GHz. At the redshift of Himiko, the \cotwo\ line ($\nu_{rest} = 230.538$~GHz) 
is redshifted to 30.354~GHz. The GBT beam at this frequency has a 
full-width-at-half-maximum (FWHM) of $\sim 25''$, corresponding to a spatial extent 
of $\sim 140$~kpc at $z=6.6$.  The GBT AutoCorrelation Spectrometer was set up to 
sample two 800~MHz IF bands at this frequency, overlapping by 200~MHz, with a single
polarization and 2048 channels per band. The total velocity coverage was $\sim 
14000$~\kms. Bandpass calibration was carried out using sub-reflector nodding with 
a cycle time of 6~seconds. While sub-reflector nodding has high observing overheads, 
it has been shown to yield excellent spectral baselines in earlier GBT Ka-band 
observations, far better than have been obtained with standard position-switching. 
Observations of 3C147 were used to correct the large-scale baseline structure in the 
target spectra and to calibrate the flux density scale, which we estimate to be 
accurate to 10\%. Typical system temperatures were 35~K throughout the observations. 
The total on-source time was 4~hours.


The data were analyzed using the {\sc GBTIDL}\footnote{http://gbtidl.sourceforge.net} 
data analysis package, following standard procedures. The data were first visually 
inspected and any data affected by correlator failures or radio frequency interference 
were edited out. Next, the narrow expected line width (250~km/s, or 25.5~MHz if the CO linewidth is similar to that of the Ly$\alpha$ line) relative 
to the size of each 800~MHz IF band, meant that we were able to fit and subtract out 
a 5th-order polynomial baseline from each individual spectrum, without affecting 
structure on the scales of the expected line emission. After flux density and bandpass 
calibration and baseline subtraction, the data were inspected for any residual 
baseline structure that might introduce spurious signals into the final spectrum. 
Around 10\% of the data were found to be affected by such artefacts and were edited 
out. The individual spectral records were than averaged together, weighting by the 
system temperature, to produce the final CO spectrum. This was then Hanning-smoothed
and re-sampled, before further boxcar smoothing and re-sampling to resolutions of 
50, 100, 150 and 250~km/s.

\section{Results and Discussion}

\begin{figure*}
\centering
\includegraphics[scale=0.45]{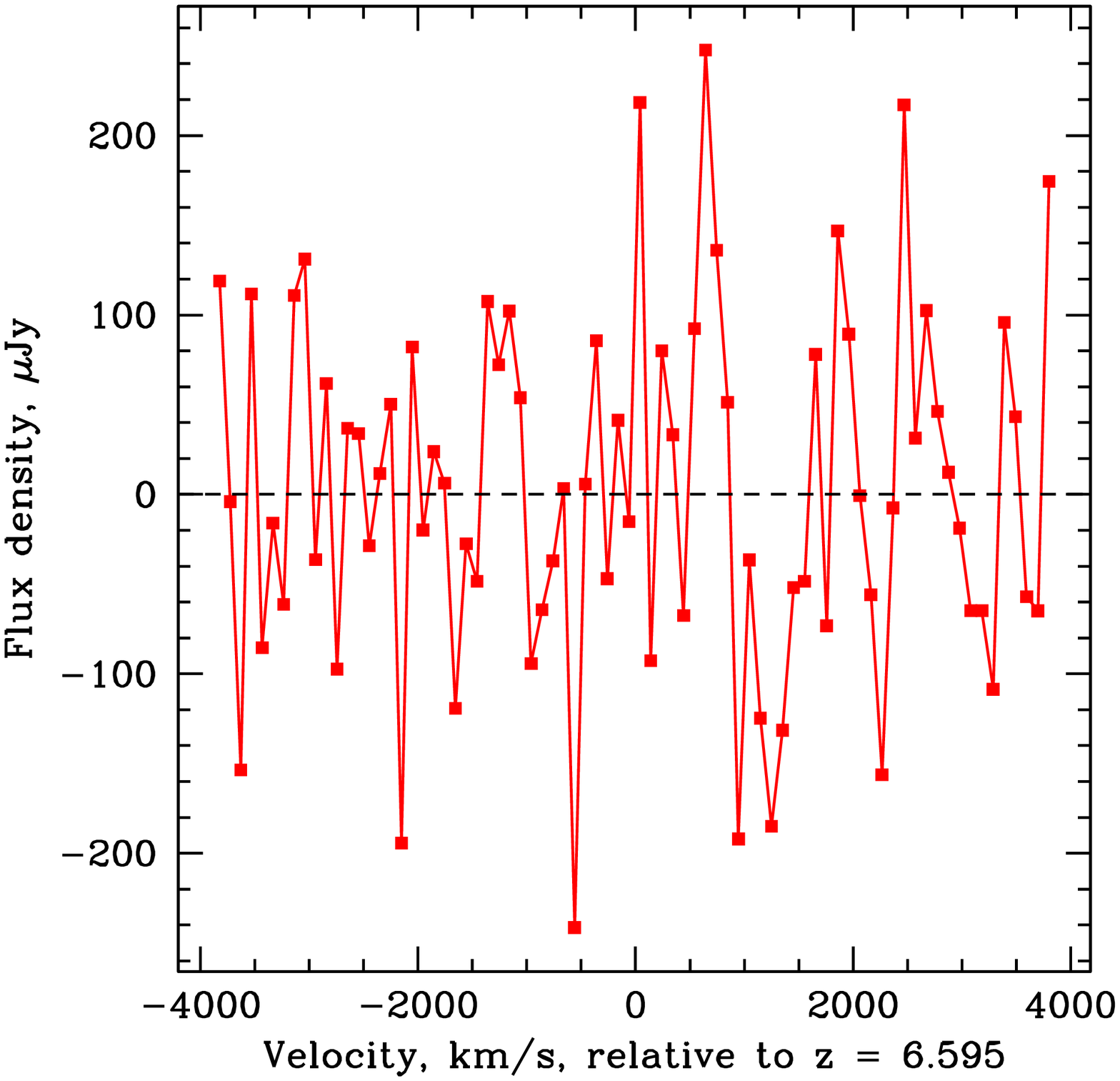}
\includegraphics[scale=0.45]{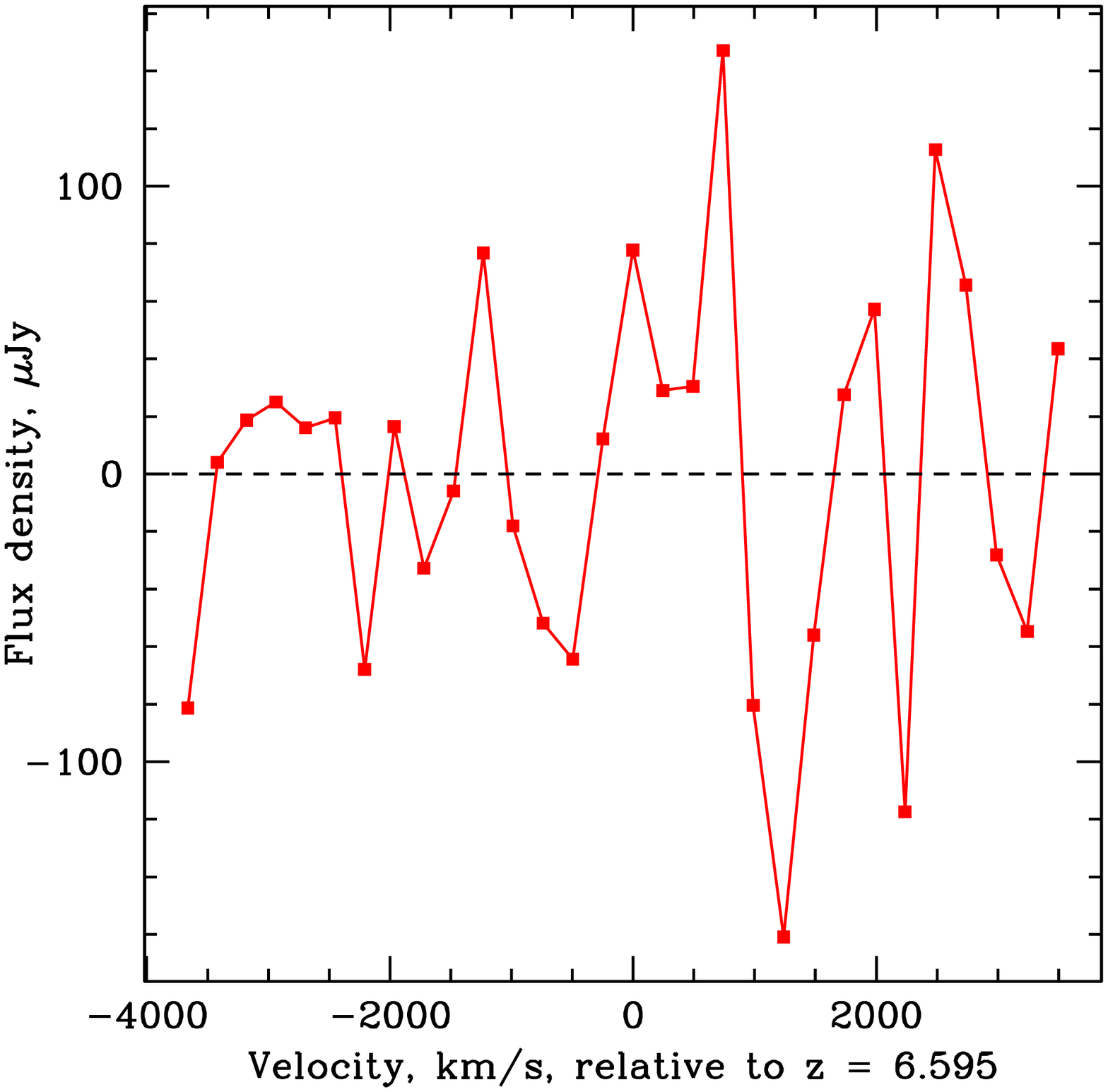}
\caption{Final GBT Ka-band spectra from the $z \sim 6.595$ LAB, Himiko, covering 
the expected redshifted frequency of the \cotwo\ line emission; the velocity scale 
is relative to $z = 6.595$. The spectra in the left and right panels have been 
smoothed to, and re-sampled at, velocity resolutions of 100~\kms\ and 250 \kms, 
respectively. The RMS noise on the spectra are $99 \mu$Jy per 100~\kms\ channel 
and $68 \mu$Jy per 250~\kms\ channel, respectively.}
\label{Fig1}%
\end{figure*}

    
The two panels of Figure~1 show the final \cotwo\ spectra of Himiko at 
velocity resolutions of 100~\kms\ (left panel) and 250~\kms\ (right panel).
The root-mean-square (RMS) noise is 99~$\mu$Jy per 100~\kms\ channel and 
68~$\mu$Jy per 250~km/s channel. No statistically-significant spectral features
are visible in either spectrum.

Our non-detection of \cotwo\ line emission in Himiko places strong constraints on 
the CO line luminosity.  The CO line luminosity, $L^\prime_{\rm CO}$, can be written 
as 
\begin{equation}
L^\prime_{\rm CO} = 3.25 \times 10^7 \: S_{\rm CO} \: \Delta V \: \nu^{-2}_{\rm obs}\:  D_L^2 \: (1+z)^{-3} \:\: ,
\end{equation}
where $\nu_{\rm obs}$ is the observing frequency (in GHz), $D_L$ is the luminosity 
distance (in Mpc), and $L^\prime_{\rm CO}$ is in K~\kms~pc$^{2}$ \citep{solomon97}.
In the case of a non-detection, $S_{\rm CO} \times \Delta V \equiv 3\sigma_{\Delta V} 
\times \Delta V$ (in Jy~\kms) gives the $3\sigma$ upper limit on the integrated flux 
density in the \co\ line, where $\sigma_{\Delta V}$ is the RMS noise at the velocity 
resolution $\Delta V$. We assume a line width of $\Delta V = 250$~\kms, from the 
observed Ly$\alpha$ line width and similar to the median observed line 
width in quasar host galaxies at intermediate redshifts \citep[e.g.][]{carilli06}.
This results in a $3\sigma$ limit of $L^\prime_{\rm CO} < 1.8 \times 10^{10} 
\times (\Delta V/250)^{1/2}$~K~\kms~pc$^2$ on the \cotwo\ line luminosity.

Assuming that the \cotwo\ line emission is thermalized, we can use the CO line 
luminosity limit to obtain an upper limit to the cold molecular gas mass in Himiko, 
assuming an appropriate conversion factor, $X_{CO}$, from line luminosity to 
molecular gas mass \citep{solomon05}. This factor can vary between 0.8 and 
4.6~M$_{\odot}$~(K~km~s$^{-1}$~pc$^2$)$^{-1}$, depending on whether the galaxy is 
an ultraluminous infrared galaxy (ULIRG) undergoing starburst activity
\citep[$X_{CO} \sim 0.8$~M$_\odot$~(K~\kms~pc$^{-2}$)$^{-1}$;][]{downes98}, or 
a more quiescent 
object like the Milky Way \citep[$X_{CO} \sim 4.6$~(K~\kms~pc$^{-2}$)$^{-1}$; 
e.g.][]{solomon91}. This factor may be even higher in regions of low metallicity gas 
\citep[e.g.][]{hughes10,leroy11}, such as those expected to be found within the 
interstellar medium of LAEs in the early Universe. However, massive galaxies like 
Himiko are unlikely to have such low metallicities, especially given its high inferred 
stellar mass ($> 10^{10}$~M$_\odot$).

Previous studies of \co line emission in two LAEs at $z\sim 6.5$ by \citet{wagg09} 
adopted a low $X_{CO}$ factor, similar to ULIRG values, based on the elevated 
star formation rates in these objects. We note that \citet{combes10} argues that 
the $X_{CO}$ factor in these systems should be closer to the Milky Way value. For 
a ULIRG conversion factor, $X_{CO} = 0.8$~M$_\odot$~(K~\kms~pc$^{-2}$)$^{-1}$, 
we obtain the limit M(H$_2$)~$< 1.4\times 10^{10}$~M$_{\odot}$ (3-$\sigma$) on the 
molecular gas mass of Himiko. \citet{ouchi09} obtain a stellar mass of 
$\sim 3.5^{+1.5}_{-2.6} \times 10^{10}$~M$_\odot$ from a fit to the spectral energy 
distribution; the molecular gas mass is thus less than the stellar mass by a 
factor of $\gtrsim 2.5$. Conversely, assuming a Milky Way conversion factor, 
$X_{CO} = 4.6$~M$_\odot$~(K~\kms~pc$^{-2}$)$^{-1}$, yields the limit 
M(H$_2$)~$ < 8.3 \times 10^{10}$~M$_\odot$ ($3\sigma$) on the gas mass.

Our limit on the CO line luminosity in Himiko can be converted to a limit on the 
total far-infrared (FIR) luminosity due to star formation . The best-fit 
parametrization of the relationship between $L_{\rm FIR}$ and $L^\prime_{\rm CO}$, 
valid for nearby spiral galaxies and starbursts, is $L_{\rm FIR} = (1.26 \pm 0.05) 
\times {\rm log} L^\prime_{\rm CO} - 0.81$ \citep{gao04}. Assuming that this 
relationship also holds 
for the interstellar medium of high-$z$ LABs, our upper limit to the CO line luminosity 
yields the upper limit $L_{\rm FIR} < 1.3 \times 10^{12}$~L$_{\odot}$ on the total 
FIR luminosity. Using the 
relation of \citet{kennicutt98b} to convert this limit on the FIR luminosity to a 
limit on the obscured SFR, we obtain $SFR < 225$~M$_\odot$~yr$^{-1}$. This is consistent 
with the lower limit of SFR~$> 36$~M$_\odot$~yr$^{-1}$ obtained from the Lyman-$\alpha$ 
line luminosity, even if obscuration/absorption effects cause the SFR derived from 
the Lyman-$\alpha$ line to be underestimated by a factor of a few.

Searches for CO emission have been carried out in only two LABs, LABd05 at 
$z \sim 2.656$ and SSA22-LAB01 at $z \sim 3.09$, prior to this work \citep{yang12}.
These authors used the IRAM~30\,m telescope to observe the CO~5-4 and CO~3-2 
transitions in LABd05 and the Plateau de Bure interferometer to observe 
the CO~4-3 and CO~3-2 lines in SSA22-LAB01. The non-detections yielded 
limits of $L^\prime_{\rm CO(5-4)} < 1.4 \times 10^{10} (\Delta V/400)^{1/2}$~K~\kms~pc$^2$ 
(LABd05) and $L^\prime_{\rm CO(3-2)} < 1.5 \times 10^{10} (\Delta V/400)^{1/2}$~K~\kms~pc$^2$ 
(SSA22-LAB01), similar to our limit on the \cotwo\ line luminosity in the 
$z \sim 6.595$ LAB. We note, in passing, that similar limits 
on the CO line luminosity have been obtained from GBT CO~{\it J}=1-0 studies of two LAEs 
at $z \sim 6.56$ and $z \sim 6.96$ \citep{wagg09}. Thus, at present, there is no 
evidence for significant amounts of molecular gas in high-$z$ LABs, despite their 
large stellar masses (e.g. \citealp{ouchi09,colbert11}). The CO luminosity in LABs 
appears to be significantly lower than that in the luminous quasar host galaxy 
population at $z > 6$ \citep[e.g.][]{walter03,wang10,wang11}, in sub-mm starburst 
and quasar host galaxies at $z \sim 2$, as well as in quiescent B$z$K-selected disk 
galaxies at $z \sim 1.5$ \citep[e.g.][]{solomon05,daddi08,aravena10}. 

Finally, it should be emphasized that the conversion factor from CO line luminosity to 
molecular gas mass is highly uncertain in high-$z$ galaxies. A low conversion 
factor [$X_{CO} = 0.8$~M$_\odot$~(K~\kms~pc$^{2}$)$^{-1}$], applicable to the 
local ULIRG population, is usually assumed for FIR-luminous objects like high-$z$ 
sub-mm galaxies and quasars \citep[e.g.][]{tacconi06}. Conversely, values closer to 
ones found in the Milky Way have been obtained in star-forming B$z$K-selected galaxies 
at $z \sim 1.5$ [$X_{CO} = 3.6\pm 0.8$~M$_{\odot}$~(K~km~s$^{-1}$~pc$^2$)$^{-2}$; 
\citealp{daddi10b}]. While studies of the high-$z$ LAB and LAE population have 
typically assumed the ULIRG conversion factor (e.g. \citealp{wagg09,yang12}), 
it is quite possible that a high conversion factor is applicable to LAEs and 
LABs at $z > 6.5$. This may be especially critical for the $z > 6$ LAE 
population, which appears to consist of relatively small galaxies with low SFRs,
$\lesssim 10$~M$_\odot$~yr$^{-1}$, and low stellar masses, $\lesssim 10^9$~M$_\odot$
(e.g. \citealp{finkelstein09a,pentericci09,ono10}). A low CO-to-H$_2$ 
conversion factor value would imply that detections of \loj\ CO line emission would 
be extremely challenging in these systems, even with new facilities like the 
Expanded Very Large Array (EVLA) or the Atacama Large Millimeter Array (ALMA). 
For example, the \cothree\ line from a $z = 6.5$ LAE with SFR=$10$~M$_\odot$~yr$^{-1}$ 
would be redshifted to a frequency of $\sim 46$~GHz with a peak flux density of 
$S_{CO} \sim 27$~$\mu$Jy, assuming a line FWHM of $\Delta V = 300$~km~s$^{-1}$), 
if we assume $X_{CO} = 4.6$~M$_{\odot}$~(K~km~s$^{-1}$~pc$^2$)$^{-1}$. Detecting 
such a line would require hundreds of hours with either the EVLA Q-band receivers 
or the ALMA Band-1 receivers. This issue would be exacerbated for the $z > 6$ 
LAEs if, as it is likely (e.g. \citealp{finkelstein11}), they are low-metallicity systems.
For such systems, with metallicities similar to the SMC, the CO-to-H$_2$ conversion 
factor could be as high as 30~M$_\odot$~(K~\kms~pc$^2$)$^{-1}$ 
\citep{hughes10,leroy11}, making it even harder to detect redshifted CO emission. 
Similarly, the models of \citet{vallini12} also suggest that the detection of \hij\ 
CO line emission in $z \sim 6.6$ LAEs will not be possible in reasonable ALMA 
integration times.

\section{Summary}

We present results from a deep GBT search for redshifted \cotwo\ line emission in 
Himiko, a Lyman-$\alpha$ blob at $z \sim 6.595$. We do not detect \cotwo\ line 
emission, with an upper limit of $L^\prime_{\rm CO}  < 1.8 \times 10^{10} \times 
(\Delta V/250)^{1/2}$~K~\kms~pc$^2$. This limit is similar to the limits on the CO 
line luminosity obtained in two LABs at $z \sim 3$ by \citet{yang12} and in two 
Lyman-$\alpha$ emitters at $z \gtrsim 6.5$ by \citet{wagg09}. Despite their high 
stellar masses, high-$z$ LABs appear to have significantly lower CO line luminosities 
than luminous quasars or sub-mm galaxies at similar redshifts. Our constraint on the 
CO line luminosity implies an upper limit of M(H$_2$)~$< 1.4 \times 10^{10}$~M$_\odot$ 
on the molecular gas mass, assuming a ULIRG conversion factor, 
$X_{CO} = 0.8$~M$_\odot$~(K~\kms~pc$^2$)$^{-1}$. However, it should be noted that 
this conversion factor is unknown for LABs. A high value of $X_{CO}$, similar to 
values in the Milky Way or $z \sim 1.5$ B$z$K galaxies, would make it extremely 
difficult to detect CO line emission from LABs and LAEs at $z \gtrsim 6$, even 
for upcoming telescopes like ALMA and the EVLA.

\acknowledgements
We thank the NRAO-Green Bank staff for their support of these observations. The 
National Radio Astronomy Observatory is operated by Associated Universities, Inc, 
under cooperative agreement with the National Science Foundation. This work was 
co-funded under the Marie Curie Actions of the European Commission (FP7-COFUND). 
NK acknowledges support from the Department of Science and Technology via a 
Ramanujan Fellowship. This work was completed during a visit by NK to ESO-Santiago; 
NK thanks ESO for hospitality and especially Alain Smette for much help during the visit.

\end{document}